# Electron-doped magnetic Weyl semimetal Li$_x$Co$_3$Sn$_2$S$_2$ by bulk-gating

Hideki Matsuoka[1, *, ¶], Yukako Fujishiro[1, *, ¶], Susumu Minami[2], Takashi Koretsune[3], Ryotaro Arita[1, 4], Yoshinori Tokura[1, 5], Yoshihiro Iwasa[1, 5]

[1] RIKEN Center for Emergent Matter Science (CEMS), Saitama 351-0198, Japan.

[2] Department of Mechanical Engineering and Sciences, Kyoto University, Kyoto 615-8246, Japan

[3] Department of Physics, Tohoku University, Miyagi 980-8578, Japan

[4] Research Center for Advanced Science and Technology, The University of Tokyo, Tokyo 153-8904, Japan

[5] Quantum-phase Electronics Center and Department of Applied Physics, the University of Tokyo, Tokyo 113-8656, Japan.

**Manipulating carrier density through gate effects, both in electrostatic charge storage and electrochemical intercalation mode, offers powerful control over material properties, although commonly restricted to ultra-thin films or van der Waals materials. Here we demonstrate the application of gate-driven carrier modulation in the microdevice of magnetic Weyl semimetal Co$_3$Sn$_2$S$_2$, fabricated from a bulk single crystal via focused ion beam (FIB). We discovered a Li-intercalated phase Li$_x$Co$_3$Sn$_2$S$_2$ featuring electron doping exceeding $5*10^{21}$ cm$^{-3}$, resulting in the Fermi energy shift of 200 meV. The carrier density dependent anomalous Hall conductivity shows fair agreement with density functional theory (DFT) calculation, which also predicts intercalated Li$^+$ ion stabilization within the anion layer while maintaining the kagome-lattice intact. This likely explains the observed rigid band behavior and constant Curie temperature, contrasting with**

**magnetic site substitution experiments. Our findings suggest ionic gating on FIB devices broadens the scope of gate-tuning in quantum materials.**

**Main**

Gate-controlled carrier modulation has been a fundamental and powerful technique in materials science research[1,2]. In condensed matter physics, for instance, gate-controlled carrier modulation has played a significant role in discovering and controlling superconductivity and magnetism, as exemplified by the study of high-$T_c$ superconductors[3,4], topological insulators[5], magnetic semiconductors[6,7], and twisted bilayer graphene[8,9]. The gating process is basically classified into two modes: electrostatic mode and intercalation mode. The former allows the charge doping only within an electrostatic screening length of typically 1 – 10 nm (ref. 1), while the latter is much larger than the screening length but cannot be in the bulk scale due to the limitation of diffusion of intercalated ions. Thus, the gating technique has been applied to ultrathin and high-quality materials systems, i.e. thin films or exfoliated flakes with nanometer-scale thickness. Consequently, suitable materials for gating are largely limited because not many materials can be prepared in the form of high-quality thin films or flakes.

In this work, we propose a new method, bulk-gating, toward tuning carriers in versatile bulk samples by combination of focused ion beam (FIB) and ionic gating. FIB presents an attractive top-down approach for fabrication of microdevices, as it enables to carve out micrometer-scale lamella from a single crystal with arbitrary crystalline orientation[10,11]. On the other hand, ionic gating, a method that controls the carrier concentration of microdevices, offers the advantage of regulating a greater number of carriers compared to conventional solid gates[12]. While one previous research applied the electrostatic mode of ionic gating to FIB device[13], this approach controlled only the surface region of the device. Here we adopted intercalation with ionic gating which enables the carrier density control not only for the surface region but also for the entire crystals by the intercalation of Li or hydrogen in the microscale devices[14–17]. Ionic gating of FIB devices should enable us to control the carrier density of the whole single crystals in the microdevices. This would broaden the materials scope for the gate-tunable devices.

In order to demonstrate the feasibility of the above concept, we have chosen $Co_3Sn_2S_2$, which is one of the most studied ferromagnetic Weyl semimetals. The crystal structure of $Co_3Sn_2S_2$ is depicted in Figure 1a and 1b, consisting of layered kagome lattice planes formed by Co and Sn atoms, and anion layers composed of Sn and S atoms[18]. One of the most notable properties of this compound is the emergence of large anomalous Hall conductivity ($\sigma_{AH}$) and anomalous Hall angle ($\theta_{AH}$), which originate from the ferromagnetism and the momentum space Berry curvature associated with Weyl nodes near the Fermi level ($E_F$)[19–21]. As demonstrated by chemical doping studies in bulk samples[22–28], the anomalous Hall conductivity as well as carrier density can be controlled dramatically by $E_F$ tuning in $Co_3Sn_2S_2$, making it an attractive test case for potential ionic gating of uncleavable materials.

**Gating device fabrication**

In this section, we summarize detailed steps for FIB device fabrication and ionic gating, as illustrated in Figure 1e. The fabrication process consists of nine steps:

- Step 1-2: Fabricate a micro-sized lamella by cutting the single crystalline bulk sample of $Co_3Sn_2S_2$ and extract it from the bulk crystal with a micromanipulator.
- Step 3: Transfer the lamella onto $SiO_2$ substrate with Au/Ti electrodes by using the micromanipulator.

- Step 4: Deposit tungsten (W) to make electrical connections between the sample and Au/Ti electrodes.
- Step 5: Remove the edges of the microdevice and make it to a Hall-bar shape by injecting ion beam. This etching process removes the thin layer of W on the side surface that is unintendedly formed during the W deposition in Step 4. We discovered that this extended W layer prevents the Li-ion insertion, which may be because the Li-ion intercalation occurs from the sample side surface, in this compound.
- Step 6: Apply a ring-shaped silicone gel to reduce the volume of liquid electrolyte placed on the device, preventing damage during cooling when the ionic liquid solidifies.
- Step 7: A Pt plate was placed above the channel region as a gate electrode, and a droplet of the electrolyte was inserted between the channel and Pt plate just before the measurements. Electrolyte was made by dissolving $LiClO_4$ (Sigma Aldrich) in polyethylene glycol PEG ($M_w$ = 600, Wako) at 80°C under vacuum beforehand to remove moisture. This step also prevents device damage caused by freezing water content. The ratio of Li to O in the PEG was set to 1:20.
- Step 8: Connect a Pt plate to the liquid and gate and apply gate voltage ($V_G$) between the gate and source. Drain voltage ($V_D$) is applied between the drain and source. During the application of $V_G$, the value of $V_D$ was set to be below 0.1V. This was done to ensure the uniform potential distribution between the drain and source regions, allowing a uniform gating effect across the sample.
- Step 9: At 330 K and under high vacuum conditions ($P < 10^{-4}$ Torr), apply a positive $V_G$. When $V_G$ exceeds a certain value, Li ions are inserted into the sample, doping electron carriers.

Figure 1c shows the scanning electron microscope (SEM) image of the fabricated device with a thickness of 1 μm at step 5, while Fig. 1d is the optical micrograph of the substrate at Step 6. The device encompasses source and drain terminals, in addition to the pair of supplementary terminals on each flank, facilitating concurrent assessment of longitudinal resistivity ($\rho_{xx}$) and transverse resistance ($\rho_{yx}$). As shown in the inset of Figure 2a, residual resistivity ratio (RRR) value, as well as the temperature dependence of longitudinal resistivity ($\rho_{xx}$-$T$), are identical between the bulk sample and the FIB device. This ensures that the bulk properties of $Co_3Sn_2S_2$ have remained unchanged in the FIB device.

**Carrier-density-dependent magneto-transport properties in $Li_xCo_3Sn_2S_2$**

Figure 2a shows a temperature-dependence of longitudinal resistivity ($\rho_{xx}$-$T$) for the FIB device at several gate voltages ($V_G$). First, we increased $V_G$ from 0 V to 4.2 V and observed a considerable increase of $\rho_{xx}$, suggesting a distinct change in electronic properties. The increase in residual resistivity is generally attributed to temperature-independent electron scattering mechanisms, such as lattice disorder and impurity. Thus, the enhancement of residual resistivity by ion gating is potentially attributed to enhanced lattice disorder by Li-ion incorporation. In contrast to the resistivity change, Curie temperature ($T_C$), corresponding to the kink anomaly in $\rho_{xx}$-$T$ curves, has shown only a small change by ion gating; $T_C$ is 170 K at $V_G$ = 0 V and 168 K at $V_G$ = 4.2 V.

After the application of $V_G$, we confirmed this Li intercalation process was semi-reversible. Figure 2a includes the $\rho_{xx}$-$T$ curve after setting the gate voltage back to 0 V after the Li intercalation and held it at 330K for 2 hours. The $\rho_{xx}$-$T$ curve after this process shows the intermediate value between the curves at $V_G$ = 0 V (before gating) and $V_G$ = 4.2

V. In general, there are various electrochemical reactions that affects $\rho_{xx}$ including removal of Sn and S elements, but these reactions are expected to be irreversible and hence the observed reversibility in the present experiment suggests that Li-ion intercalation was the most likely process. On the other hand, the fact that the pristine state is not completely recovered even after setting $V_G$ back to 0 V suggests that the Li intercalated $Co_3Sn_2S_2$ is rather stable. This semi-reversibility was confirmed in terms of the anomalous Hall effect, too, as discussed later.

Figure 2b shows the Hall conductivity ($\sigma_{yx} = \rho_{yx} / (\rho_{yx}^2 + \rho_{xx}^2)$), as a function of out-of-plane magnetic field ($\mu_0 H$) at $T = 10$ K. The signal can be decomposed into two terms $\sigma_{yx} = \sigma_{AH}(M_z) + \sigma_n(\mu_0 H)$, with the former term corresponding to the anomalous Hall effect with a hysteresis behavior and the latter term associated with the normal (Lorentz-force) Hall effect. First, we discuss the gating effect on the normal Hall effect, focusing on the variation of the carrier density. A noticeable feature is the sign change of dominant carrier by gating, as manifested in the slope of the normal Hall resistivity. Upon applying $V_G = 4.2$ V, the $B$-linear term with a negative slope is detected (see the inset of Fig. 2b), corresponding to the electron carrier concentration of $n_e \sim 6.2 \times 10^{21}$ cm$^{-3}$. This value is a few orders of magnitude greater than the that observed in pristine compensated semimetal $Co_3Sn_2S_2$ ($n_h \sim 9.3 \times 10^{19}$ cm$^{-3}$ and $n_e \sim 7.5 \times 10^{19}$ cm$^{-3}$)[19]. If we assume that Li ion releases less than one electron, the observed large change in electron carrier density corresponds to the chemical composition of Li$_x$Co$_3$Sn$_2$S$_2$ with $x$ value being larger than 0.6. By setting the gate voltage to 0 V after the gate application, the electron carrier density was reduced. The carrier concentration at $V_G = 0$ V (after gating) was returned to $n_e = 1.5 \times 10^{21}$ cm$^{-3}$, which corresponds to the de-intercalation of Li ions resulting on the change of $x$ in Li$_x$Co$_3$Sn$_2$S$_2$ from 0.70 (at $V_G = 4.2$ V) to 0.17 (at $V_G = 0$ V, after gating).

Next, we focus on the gate dependence of the anomalous Hall effect at zero magnetic field. Firstly, modulation of the coercive field ($H_c$) is observed in all anomalous Hall effect data. Each dataset represents a single, not multi-step, square signal, indicating a uniform phase throughout the device. This confirms that the entirety of the μm-sized device is being uniformly gate-modulated in the combination of the FIB device and the ion gate. Subsequently, we discuss the gate modulation of the magnitude of the anomalous Hall conductivity. A substantial modulation of $\sigma_{AH}$ from 960 $\Omega^{-1}$cm$^{-1}$ ($V_G = 0$ V, $T = 10$ K) to 290 $\Omega^{-1}$cm$^{-1}$ ($V_G = 4.2$ V, $T = 10$ K) was observed by gating. We also note that while the gating suppresses the anomalous Hall conductivity, the anomalous Hall angle ($\theta_{AH} = \sigma_{AH} / \sigma_{xx}$) remains high around $\theta_{AH} = 0.11$ ($V_G = 4.2$ V, $T = 10$ K), suggesting a large contribution from the Berry curvature stemming from the Weyl nodes. This may imply that the electronic and magnetic structures, which are largely dictated by the kagome-lattice geometry, are not significantly altered by the ion gating. Indeed, we discovered that the observed changes in carrier concentration and anomalous Hall conductivity shows striking agreement with the DFT calculations in terms of Fermi energy tuning within a rigid band model, as will be discussed later. The semi-reversible gating effect is also confirmed by the anomalous Hall conductivity, which shows $\sigma_{AH} = 640$ $\Omega^{-1}$cm$^{-1}$ at $V_G = 0$ V (after gating) as shown in Fig. 2b.

Here, we make a quantitative comparison of the experimental results with the DFT calculations. Figure 3a shows the calculated band structure of $Co_3Sn_2S_2$, revealing the presence of two sets of Weyl nodes with a finite gap positioned slightly above and below the $E_F$, respectively. Figures 3b and 3c present the $E_F$ dependence of hole ($n_h$) and electron ($n_e$) densities, as well as $\sigma_{AH}$ in $Co_3Sn_2S_2$. The crystal structure for this calculation is assumed to be identical to the pristine material, and $E_F$ was scanned following the rigid band picture. The carrier concentration after gate voltage application, $n_e = 6.2 \times 10^{21}$ ($V_G = 4.2$ V, $T = 10$ K) corresponds to the $E_F$ shift by +200 meV to the electron side. In this doping level the calculation predicts almost the same anomalous Hall coefficient as the experimentally observed $\sigma_{AH}$ = 290 $\Omega^{-1}$cm$^{-1}$ ($V_G$ = 4.2 V, $T$ = 10 K). Another thing to be noted is that $n_e$ and $\sigma_{AH}$ are highly reproducible in other devices (see supplementary information section A in details). This remarkable reproducibility and agreement with the DFT calculations[34] lead us to the following conclusions. The first is that there is a rather stable phase of Li$_x$Co$_3$Sn$_2$S$_2$ with $x$ > 0.6, and the second is that the $E_F$ shift occurs in a rigid band manner. Such a rigid band manner is also applicable after the removal of the gate voltage to $V_G$ = 0 V a, where $n_e$ and $\sigma_{AH}$ value also shows the agreement with the calculations with the $E_F$ shift by +100 meV to the electron side (Fig. 3).

**Carrier-density-dependence on Curie temperature in Li$_x$Co$_3$Sn$_2$S$_2$**

Figures 4a and 4b display the temperature dependence of the Hall angle and $\sigma_{xy}$, respectively. $T_C$ determined by the emergence of the anomalous Hall signal was found to be consistent with the kink anomaly temperature in $\rho_{xx}$-$T$ curves (supplementary information section D). The change of $T_C$ is minimal despite the considerably large carrier density modulation by two orders of magnitude. As shown in Figure 5c, $T_C$ is almost independent of carrier density for all the devices and gate voltages.

Here we compare the carrier-density-independent nature of $T_C$ discovered in Li$_x$Co$_3$Sn$_2$S$_2$ with other studies of carrier doping in Co$_3$Sn$_2$S$_2$. So far, electron carrier doping in Co$_3$Sn$_2$S$_2$ has only been achieved through chemical substitution, which can be categorized into two approaches: substitution of Co with Ni, and substitution of Sn with Sb. The relationship between the amount of chemical substitution and $T_C$ in these previous studies is shown in Figure 5a. In both cases, $T_C$ decreases, while the change is rather moderate in the case of Sb doping compared to the case of Ni doping. Hence, Li$_x$Co$_3$Sn$_2$S$_2$ presents a distinct trend, in a sense that $T_C$ is almost independent of carrier density control.

These distinct tendencies between three different approaches of carrier doping may be attributed to the chemical impurity effect on magnetism. As discussed earlier, the crystal structure of $Co_3Sn_2S_2$ is characterized as stacked kagome layer-anion layer structure, where the magnetic kagome layers are composed of positively charged $Co^{\delta+}$ cations while the anion layers are composed of $S^{2-}$ anions. The Ni doping directly replaces the magnetic Co atoms in kagome-layer, which should induce a crucial effect on $T_C$. On the other hand, Sb doping replaces some atoms in the anion layer, allowing carrier doping with less impact on the kagome lattice. In the case of $Li_xCo_3Sn_2S_2$, we expect that positively charged $Li^+$ ions are preferentially inserted into the anion layers, forming Li-S bonds while maintaining the magnetic kagome layers completely intact. The DFT calculations suggest that such Li-S bonds shown in Figure 5b and 5c indeed has a local minimum in energy. This result provides important implications for the selection of material groups that will be targeted in future studies on bulk-gating control.

It is highly likely that our study has achieved 'clean' doping through Li intercalation, and revealed that the $T_C$ of $Co_3Sn_2S_2$ is independent of carrier concentration. This is quite astonishing as it differs from the common situations of itinerant ferromagnets where $T_C$ depends on carrier concentration. For instance, the origin of magnetism in itinerant ferromagnets is usually explained by the Stoner model or RKKY interactions, both of which are strongly dependent on the electronic states near the Fermi level ($E_F$), which is contrary to the behavior observed in this study. The present observation implies the mechanism of interaction is the direct exchange between the localized magnetic moments rather than the carrier-mediated ones. On the other hand, several previous studies on magnetic semimetals also report the lack of change in magnetic transition temperature[29, 30]. To explain the origin of ferromagnetism in these systems, mechanisms such as Bloembergen–Rowland or Van Vleck have been proposed[31, 32]. Both mechanisms are linked with the narrow-gap band character typical of magnetic semimetals, potentially elucidating the behavior observed in $Co_3Sn_2S_2$.

In summary, this study successfully combined focused ion beam (FIB) device fabrication and ion-gating techniques (Li-ion intercalation) to realize gate-controlled carrier modulation to bulk microdevice. Using $Co_3Sn_2S_2$ as a target material, we achieved $E_F$ modulation as large as 200 meV, which was estimated from the changes in carrier density and anomalous Hall effect. Notably, we utilized the minimal impact of Li-ion intercalation on magnetism to uncover that the $T_C$ of $Co_3Sn_2S_2$ does not depend on doped carrier density. This unique behavior of $T_C$ is expected to provide insights into the origin of magnetism in $Co_3Sn_2S_2$. In addition, gate control of bulk material as demonstrated in this study will be highly advantageous especially when is hard to obtain ultrathin films or flakes of the target material. Here, we propose a unique layered crystal structure of $Co_3Sn_2S_2$, where the anion layers host $S^{2-}$ ions, likely allows the control of carrier density by Li intercalation. This insight is expected to dramatically expand the range of target materials for future ion gating study. For example, an extensive family of uncleavable kagome-lattice magnets and superconductors, featuring intertwined topology and electron correlation[33], could be of significant interest. The work provides a significant step forward in bridging quantum material science and gate control nanotechnology.

...

**Methods**

**Bulk sample fabrication and sample characterization.**

Single crystalline samples of $Co_3Sn_2S_2$ were prepared by the Bridgman method. First, Co, Sn, and S were mixed in a stoichiometric ratio and then sealed in a quartz tube. Afterwards, it was heated up to 1323 K and cooled down to 973 K with a rate of 4 mm per hour. The shandite-type crystal structure was confirmed by using the powder x-ray analysis.

**Transport measurements.**

The electrical transport properties of the ion-gating devices were characterized by combination of a source-measure unit (Agilent Technologies, B2912A) and voltmeters (Keithley Instruments, 2182A) at different temperatures and magnetic fields under different $V_G$ in Physical Property Measurement System (Quantum Design, PPMS). Hall resistivity data were anti-symmetrized against magnetic field to remove the longitudinal voltage contribution due to electrode misalignment. After applying each gate voltage at *T*

= 330 K under high vacuum condition (< $10^{-4}$ Torr), we waited for 30 minutes to ensure uniform Li-ion insertion in the sample. After cooling to 250 K, the sample chamber was purged with helium gas to improve the stability of temperature.

**Band structure calculation.**

The electronic structure of $LiCo_3Sn_2S_2$ was investigated using the Vienna ab initio simulation package (VASP) with DFT[36-38]. The exchange-correlation functional was described using the generalized gradient approximation of Perdew-Burke-Ernzerhof[39]. The cut off energies for a charge density of 520 eV, Gamma-centered $k$ meshes of 12x12x12 were used in the self-consistent band structure calculation. The lattice constant was set to the experimental values, and the ionic positions were obtained through all-ions relaxation.

**Data availability.**

The data within this paper are available from the corresponding author upon reasonable request.

**Acknowledgments**

We are grateful to N. Jiang and M. Tanaka for valuable discussions. This work was supported by the Grants-in-Aid for Scientific Research (Grant No. 22K14011, No. 21K13888, No. 19H05602, No. 22K14587, 21H04437, 21H04990, 19H05825), A3 Foresight Program from the Japan Society for the Promotion of Science (JSPS). This study was carried out jointly by the Cryogenic Research Center and the University of Tokyo.

**Corresponding Author**

*E-mail: hideki.matsuoka@riken.jp

*E-mail: yukako.fujishiro@riken.jp

**Author contributions**

¶H.M. and Y.F. contributed equally to this study. Y.F. grew and characterized the bulk samples and fabricated FIB devices. Y.F. performed the magnetization and transport measurements on bulk and ungated FIB devices. H.M. performed gating experiments. H.M. and Y.F. analyzed the data. S.M., T.K. and R.A. performed the first-principles calculations. Y.T. and Y.I. supervised this study. H.M., Y.F., Y.T. and Y.I. wrote the manuscript. All the authors discussed the results and commented on the manuscript.

**Additional Information**

**Supplementary Information** accompanies this paper at ---.

**Competing interests:** The authors declare no competing interests.

**Reprints and permission** information are available online at ---.

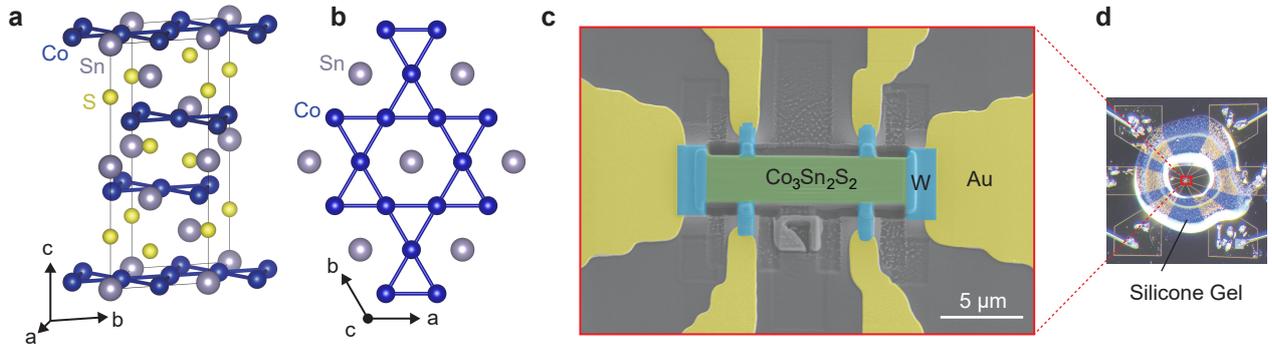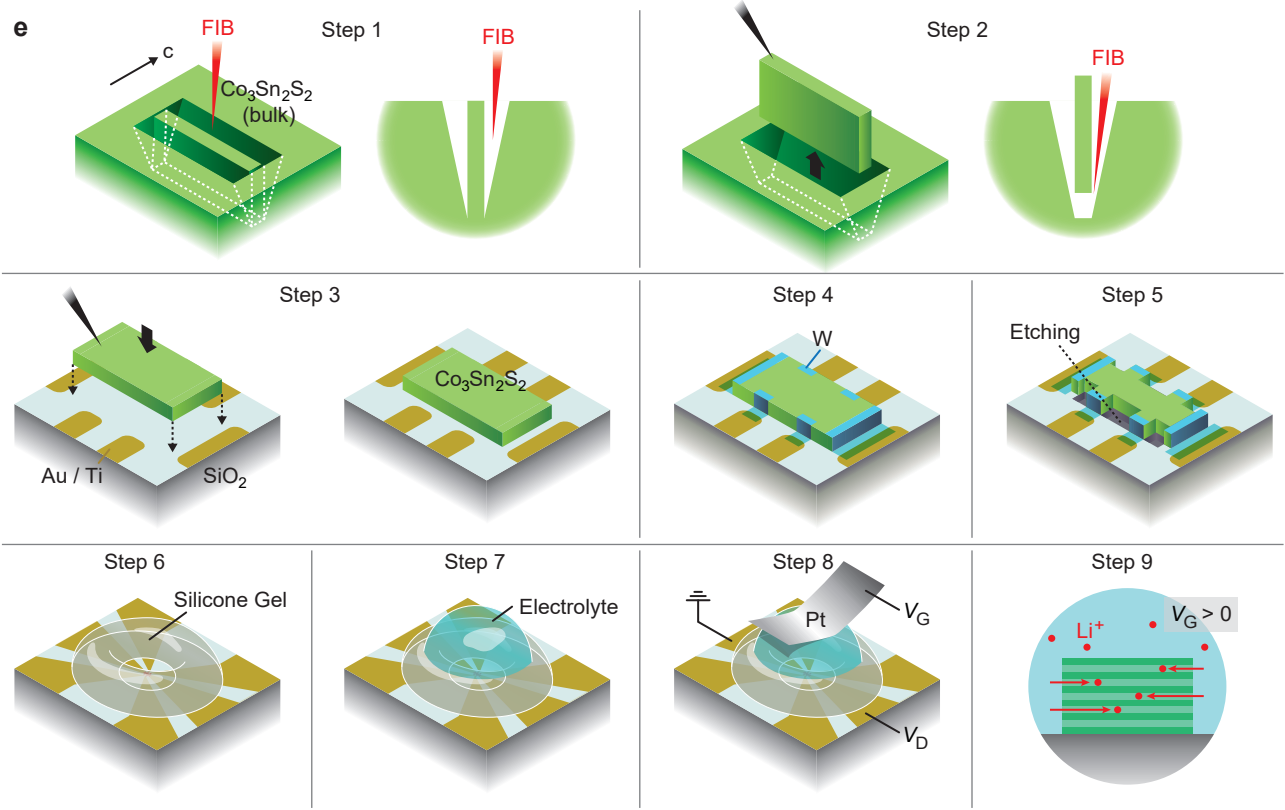

**Figure 1 | Schematic illustration of the fabrication procedure steps for $Co_3Sn_2S_2$ gating device.**

**a,b,** Top (**a**) and side (**b**) views of $Co_3Sn_2S_2$ crystal structure by VESTA[34]. **c,d,** SEM image of the FIB device with a thickness of 1 μm after step 5 (**c**) and optical microscope image for step 6 (**d**). **e,** Schematic illustrations for fabrication process.

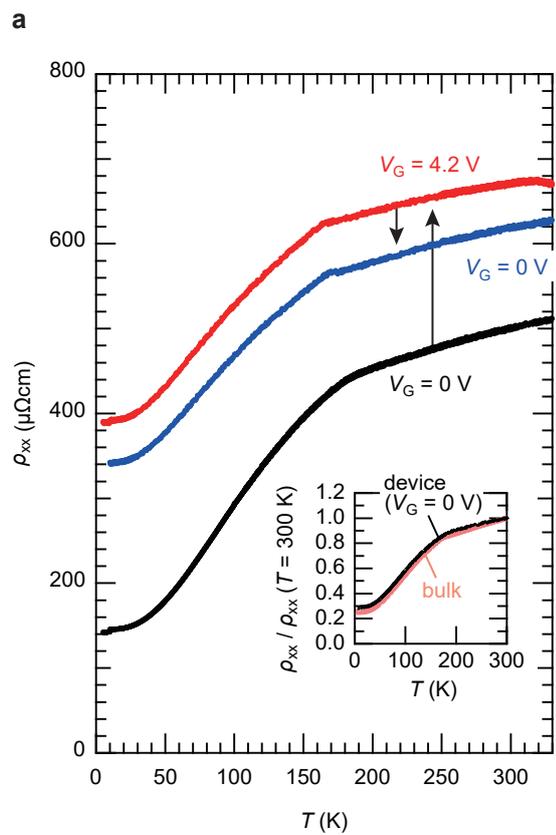 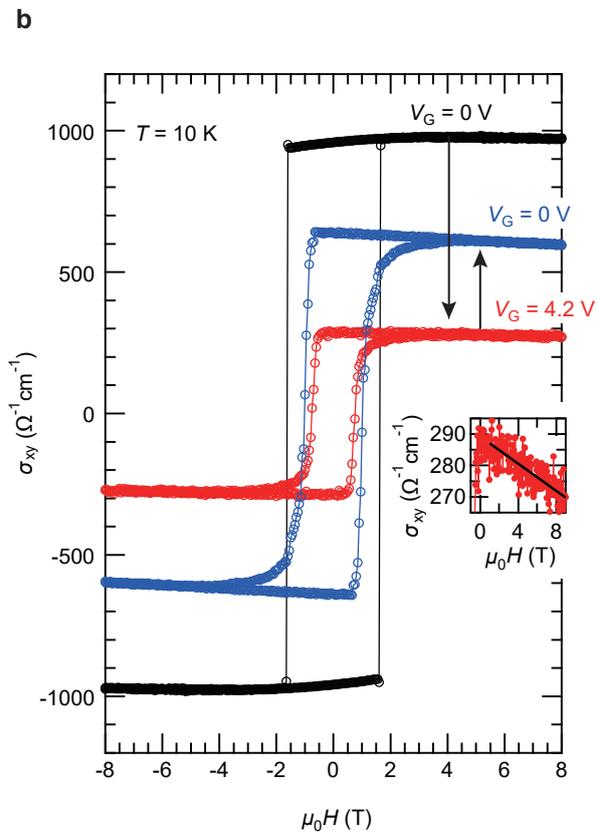

**Figure 2 | Gate voltage dependence of transport properties in $Co_3Sn_2S_2$.**

**a,** The temperature dependence of longitudinal resistivity ($\rho_{xx}$) as a function of temperature (*T*) of device C with $V_G$ = 0 V (before applying gating), 4.2 V and 0 V (after applying gating), respectively. The inset shows $\rho_{xx}$-*T* curves before and after device fabrication, being normalized by the value at *T* = 300 K for comparison. **b,** The Hall conductivity ($\sigma_{yx} = \rho_{yx} / (\rho_{yx}^2 + \rho_{xx}^2)$) versus applied magnetic field ($\mu_0 H$) of device C at *T* = 10 K with $V_G$ = 0 V (before applying gating), 4.2 V and 0 V (after applying gating), respectively. The inset shows the magnified data of $\sigma_{yx}$ ($\mu_0 H$) for $\mu_0 H$ > -1 T at $V_G$ = 4.2 V.

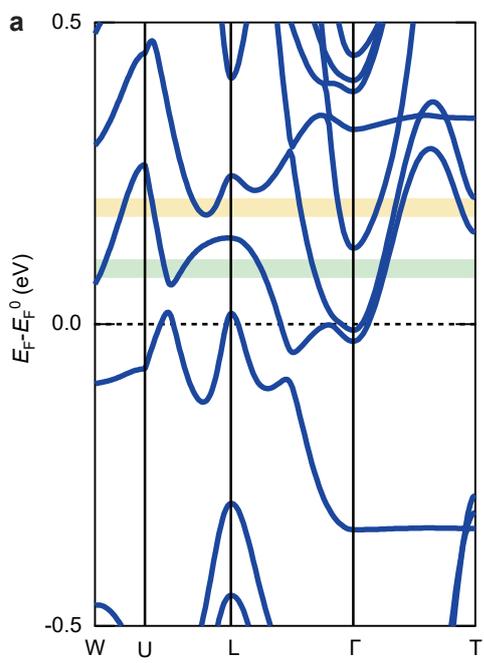
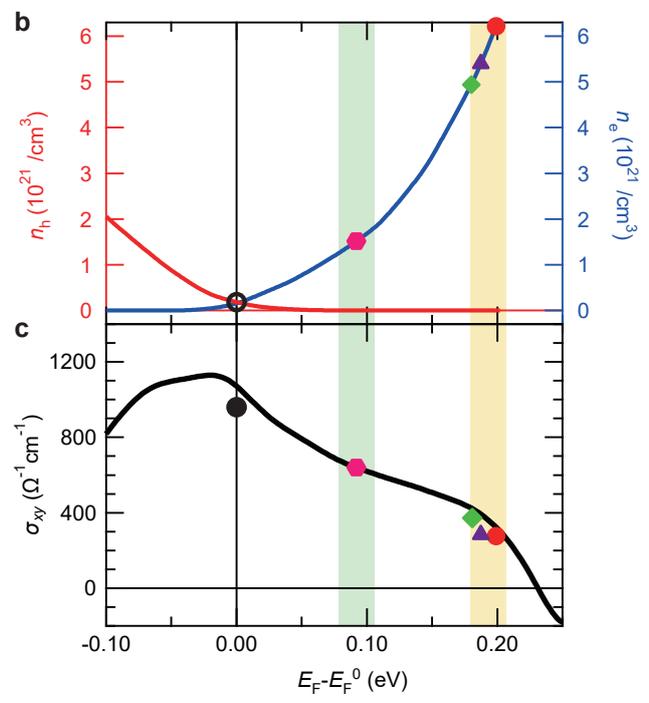

**Figure 3 | The calculated band structure, carrier density, and the anomalous Hall conductivity in Co$_3$Sn$_2$S$_2$.**

**a,** The band structure of Co$_3$Sn$_2$S$_2$ with spin-orbit coupling along the high-symmetry paths. **b,** The blue and red lines show the calculated density of electron $n_e$ and hole $n_h$, respectively. The open black circle refers the carrier density of bulk Co$_3$Sn$_2$S$_2$ taken from the previous study[19]. **c,** The calculated anomalous Hall conductivity ($\sigma_{AH}$) as a function of $E_F$. The solid symbols are the experimentally obtained values. All calculations data are reproduced from the previous paper[31].

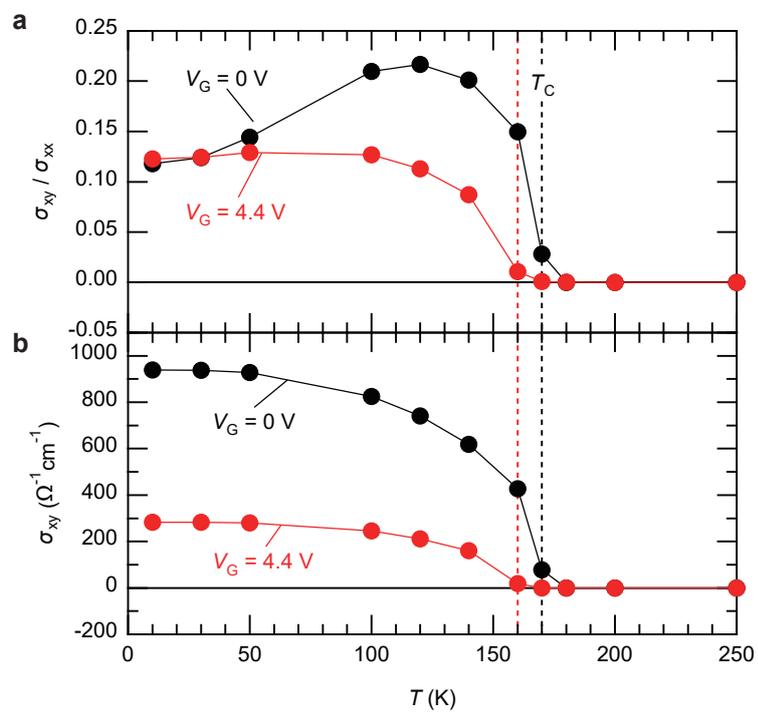

**Figure 4 | Temperature dependence of the anomalous Hall effect.**

**a,b,** Temperature dependence of anomalous Hall angle, $\theta_{AH} = \sigma_{AH} / \sigma_{xx}$, (**a**) and the anomalous Hall conductivity, $\sigma_{AH}$, (**b**) at $V_G = 0$ V and 4.4 V of device B, respectively. The dotted lines represent the Curie temperature ($T_C$) for $V_G = 0$ V (black) and 4.4 V (red), defined by the kink anomaly in $\rho_{xx}$-$T$ curves in supplementary information section D, which is consistent with the temperature where the anomalous Hall signal starts to emerge.

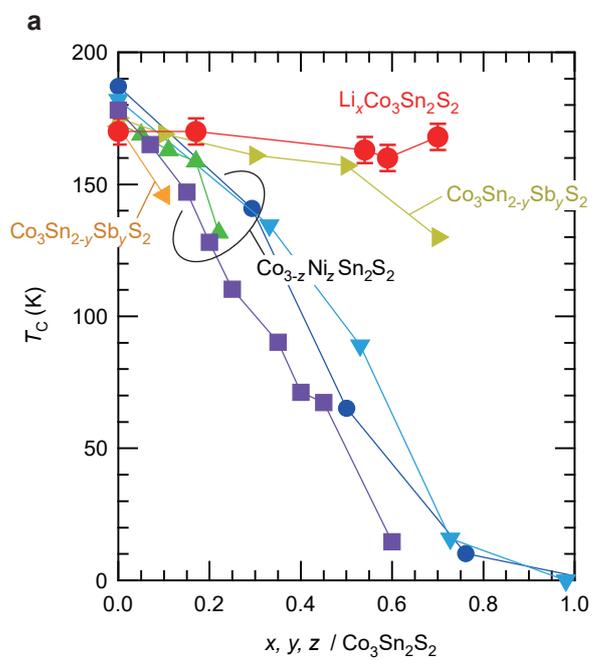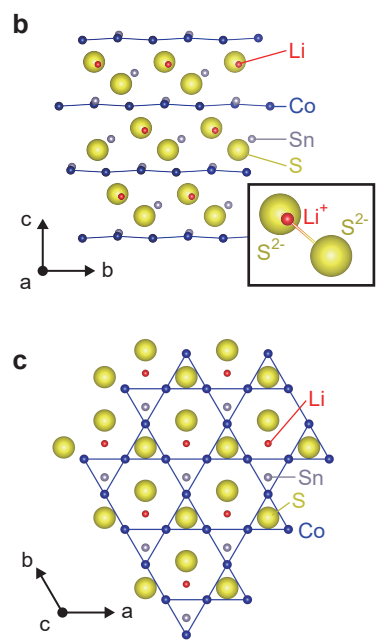

**Figure 5 | Properties of Li-ion intercalated Co$_3$Sn$_2$S$_2$, Li$_x$Co$_3$Sn$_2$S$_2$.**

**a,** Comparison of $T_C$ between chemically doped samples[22, 23, 26-28] and the gated microdevice in this study. The *x* of Li$_x$Co$_3$Sn$_2$S$_2$ was derived from the electron carrier density, assuming the density of Li$^+$ ion is equal to the doped electron carrier density. **b,c,** Top (**b**) and side (**c**) views of potential crystal structure of LiCo$_3$Sn$_2$S$_2$ drawn with VESTA[34]. Li$^+$ ion is inserted in anion layers forming Li-S bonds, as shown in the inset. The crystal structure is obtained from the DFT calculations. The size of each atom in this figure corresponds to the ionic radius, which is different from that in Fig. 1a and 1b where the size of each atom corresponds to the atomic radius.